\documentclass[sn-mathphys,Numbered]{sn-jnl}

\usepackage{graphicx}%
\usepackage{multirow}%
\usepackage{amsmath,amssymb,amsfonts}%
\usepackage{mathrsfs}%
\usepackage[title]{appendix}%
\usepackage{xcolor}%
\usepackage{textcomp}%
\usepackage{manyfoot}%
\usepackage{booktabs}%
\usepackage{algorithm}%
\usepackage{algorithmicx}%
\usepackage{algpseudocode}%
\usepackage{listings}%

\begin{document}

\title{Non-equilibrium physics of multi-species assembly: \\
From inhibition of fibrils in biomolecular condensates to growth of online distrust }


\author*{\fnm{Pedro D.} \sur{Manrique}}\email{pmanriq@gmail.com}
\author{\fnm{Frank Yingjie} \sur{Huo}}
\author{\fnm{Sara} \sur{El Oud}}
\author*{\fnm{Neil F.} \sur{Johnson}}

\affil{\orgdiv{Physics Department}, \orgname{George Washington University}, \city{Washington}, \postcode{20052}, \state{DC}, \country{U.S.A}}

\abstract{Self-assembly is a key process in living systems -- from the microscopic biological level (e.g. assembly of proteins into fibrils within biomolecular condensates in a human cell) through to the macroscopic societal level (e.g. assembly of humans into common-interest communities across online social media platforms). The components in such systems  (e.g. macromolecules, humans) are highly diverse, and so are the self-assembled structures that they form. 
However, there is no simple theory of how such structures assemble from a multi-species pool of components. Here we provide a very simple model which trades myriad  chemical and human details for a  transparent analysis, and yields results in good agreement  with recent empirical data. It reveals a new  inhibitory role for biomolecular condensates in the formation of dangerous amyloid fibrils, as well as a kinetic explanation of why so many diverse  distrust movements are now emerging across social media. The nonlinear dependencies that we uncover suggest new real-world control strategies for such multi-species assembly.}

\keywords{Nonequilibrium statistical physics, amyloid fibrils, biomolecular condensates, online distrust networks, control strategies}

\maketitle
Somehow, nature manages to continually build a rich variety of mesoscopic and macroscopic structures from the cellular to societal levels -- starting with some huge bag of heterogeneous components of different types. Undoubtedly these processes involve  considerations of temperatures, free energies, chemical compositions, surface phenonema etc. However, including all such factors in a calculation is impossible -- let alone the added features of the system being open and out-of-equilibrium -- and it misses the point of providing a simple understanding of self-assembly across living systems' scales and domains.  For example, in cancer biology a tumor microenvironment (TME) is a complex and evolving heterogeneous network containing cellular and non-cellular components, which extends beyond the heterogeneous population of tumor cells and includes a wide variety of the host’s immune cells \cite{jin20,zare20,milanowski22}. The TME components and their interactions regulate drug escape mechanisms, metastasis, the suppression of the immune system response, and other mechanisms conducive to the development and progression of the disease \cite{milanowski22,zare20,Nabet19,shevde19}. Similarly, recent experiments have demonstrated how the onset of amyloid fibrils can drastically be impacted by the properties of the biological environment where it develops \cite{arosio21,spruijt22}. Environments consisting on multi-domain biomolecular condensates have shown to delay the onset of the amyloid fibrils of the A$\beta$-42 peptide \cite{arosio21}. This is despite the enhanced local concentration of amyloidogenic proteins enabled by the biomolecular condensate, which go against the law of mass action. Acceleration, as well as, suppressed aggregation have also been observed in the amyloidogenic protein $\alpha$-synuclein while in biomolecular condensates composed on non-aggregating peptides \cite{spruijt22}. The challenge of how to better characterize the complex and dynamical interactions between a sub-system and its complex environment so that appropriate control strategies can be outlined, remains unresolved.\\

\noindent A parallel of this self-assembly occurs at the other end of the living-systems spectrum, in the seemingly unrelated field of social systems. Subsets of  diverse actors manage to somehow self-assemble into communities where they carry out coordinated activity to attain specific goals. Online social media is one such example, comprising interconnected online communities (e.g. Facebook Groups) formed by diverse individuals from anywhere across the globe, who interact in a distance-independent way around their preferred topic. These topics can be contentious, e.g., vaccine safety \cite{johnson2020online,lucia22}. Similar to the biological domain where even individuals of the same species can have phenotypic variations \cite{vasseur11,seifert15,nelson21,xue23}, online communities are highly diverse even when formed around specific contentious stances. These communities and their members, can have a wide range of beliefs, come from a wide range of backgrounds and locations, and yet manage to somehow self-assemble and hence the community can grow \cite{johnson2020online,lucia22,VKScience16,prl18,BoogSciRep21,PRL2023}.\\

\noindent The commonality of such systems lies in the huge heterogeneity of their components and interactions, even among members of the same species. Here we fill a significant gap in the literature by providing a first-principles theory explicitly accounting for this diversity both within and across species of components.  Interactions are dictated by individual objects’ affinity, for both intra- and inter-species. Our theory goes beyond the one-species analysis in Ref. \cite{prl18,PRL2023}. We purposely test it in two disparate application domains: (1) the formation of amyloid fibrils within multi-domain biomolecular condensates and (2) the growth of vaccine distrust in the social media platform Facebook. We also show how the complex nonlinear dependencies that it uncovers, suggest new control strategies in all such self-assembly systems.

\begin{figure}
\centering
\includegraphics[width=0.9\linewidth]{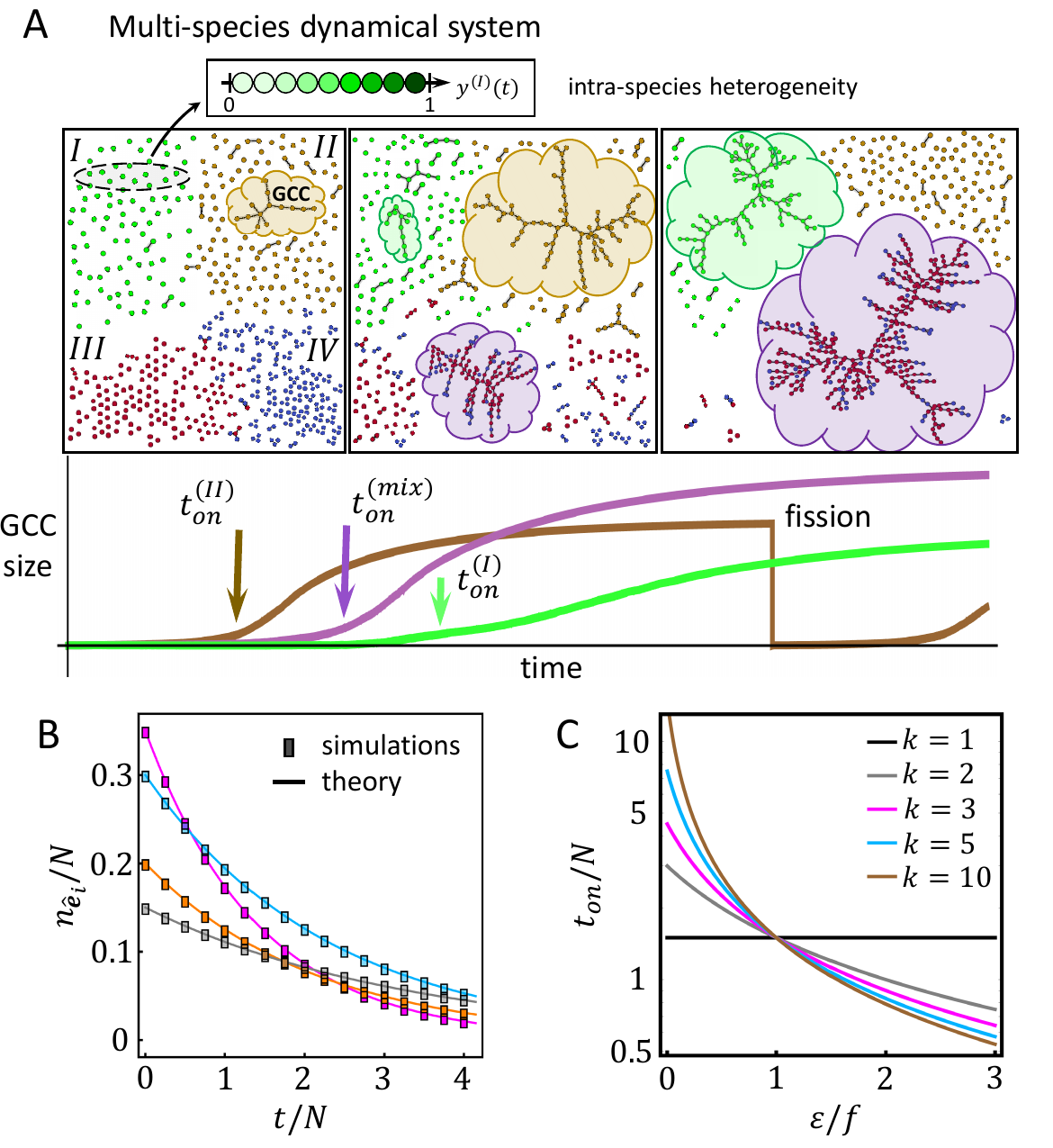}
\caption{\small{{\bf Multi-species framework.} A. Schematics of the system, interactions and dynamics of 4 co-evolving species of different sizes, interaction rates, and properties: I (green), II (brown), III (red) and IV (blue). The heterogeneity of each individual object  within its own species $j$, is represented by a function $y^{(j)}(t)$ ranging from 0 to 1. The central panel illustrates the state of the system at three specific times. Species I and II co-evolve without mixing, i.e., they form pure clusters, while species III and IV form mixed clusters. Clusters occasionally experience total fragmentation (i.e., fission). The bottom panel shows the size of the Giant Connected Component (GCC) of the pure and the mixed sub-systems pointing to large scale events such as the onset and total fission. B. Dynamics of monomers in a system of $k=4$ species as a function of time, where the theoretical results are compared with those from microscopic simulations. The parameters are: $N_1=700$ (magenta), $N_2=600$ (cyan), $N_3=400$ (orange), $N_4=300$ (gray), the elements of the F-matrix are $f_{11}=f_{44}=1$, $f_{22}=1/2$, $f_{33}=2/3$, $f_{ij}=(1/3)\delta_{2,3}$, where $\delta_{i,j}$ is the Kronecker delta with $i,j=1,...,k$ for all $i\neq j$. C. Onset time as a function of the ratio of cross-species interaction $\varepsilon$ to the same-species interaction $f$, for a system of $k$ species for the cases shown.}}
\label{fig1}
\end{figure}

\section{Multi-species Theory}
\noindent Our theory include two levels of heterogeneity -- the species level, and then heterogeneity within each species -- though this could be further generalized to even higher levels of species-of-species by re-scaling. We consider a system of $N(t)$ individual components, each of which belongs to one of $k$ species types and each of which has its own heterogeneity within this species type. These $N(t)$ components undergo fusion/fission dynamics according to specific rules dictated by the objects' mutual affinity. The individual within-species heterogeneity is an endogenous set of traits that could change over time. For notation simplicity, here we consider one trait per object represented by a real number between zero and one (Fig. \ref{fig1}A). Objects of the same species interact according to a given rule (e.g., homophily), while objects of different species may interact according to a different rule (e.g., heterophily). We assume that rules do not change over time and hence objects cannot change species either. Therefore, under a mean-field approximation, we can define a $k\times k$ fusion matrix $\mathbf{F}(t)$ quantifying the mean-field fusion probability associated with interactions between same-species objects in its diagonal entries ($f_{ii}(t)$), and between different species objects in the off-diagonal entries ($f_{ij}(t)$, $i\neq j$). This definition can be generalized to account for higher-order interactions where $\mathbf{F}(t)$ becomes a tensor of rank greater than 2. By contrast, fission processes consist of a full fragmentation of any multi-species cluster of size $c$ into $c$ clusters of size one with a probability $\nu_{\mathrm{f}}$, which for simplicity, we set it equal across species. The interplay between these processes dictates the dynamics of observables such as the number of monomers (Fig. \ref{fig1}B). To check the validity of the analytic results we employ extensive computational simulations to map the dynamics of the system at the level of individuals (see Methods). If the fission probability is very small compared to the fusion, the system could experience a large-scale connectivity transition where the largest cluster, or the giant connected component (GCC), encompasses a non-negligible fraction of a specific population (Fig. \ref{fig1}A). Critical questions surrounding this phenomenon include the impact of individual heterogeneity in the GCC onset. Likewise, in a multi-species framework, it becomes key to determine which species becomes prevalent in the GCC and the conditions that yield its dominance.\\

\noindent The GCC onset for the limit of negligible fragmentation is found through the singularities of the second moment of the size distribution \cite{RednerBook,NewmanBook,prl18,PRL2023} yielding (see Methods):
\begin{equation}
\det{\left(\mathbf{N}^{-1}-\frac{2}{N^2}\mathbf{F}t_{j}\right)}=0,\quad j=1,...,k
\label{eq:roots}
\end{equation}
\noindent where $t_j$ points to the $j$-th root of the resultant polynomial of degree $k$, and $\mathbf{N}$ is a $k \times k$ diagonal matrix of the number of objects $N_{j}$ of species $j$ ($j=1,...,k$) on each entry. In the limit where all the cross-species interaction values tend to zero, each root points to the individual GCC onset for each species, $t_{on}^{(j)}\rightarrow N^2/(2f_{j}N_j)$. Therefore, the system could have many GCC onsets happening at different points in time for the different species present (Fig. \ref{fig1}A). As the values of the cross-species interactions become non-zero, the GCC of coupled species merge and the smallest positive root in equation $\ref{eq:roots}$ becomes the onset $t_{on}$ of the coupled GCC rendering the remaining $t_j$ values nonphysical (see SI sec. 1.2.3 for examples). As $k$ increases so does the complexity of the mathematical expressions. Assuming equal values in the number of objects of each species, the same-species interactions ($f_{ii}\equiv f$), and the cross-species interactions ($f_{ij}\equiv \varepsilon$, $i\neq j$), a compact formula of the GCC onset resembling that of the single species result \cite{prl18} is found to be $t_{on}=N/2F_{\mathrm{eff}}$, where $F_{\mathrm{eff}}=[f+(k-1)\varepsilon]/k$ (see SI sec. 1.6). This formula uncovers an interplay between the relationship among the same- and cross-species interactions with the onset time for different $k$. If $f>\varepsilon$, the onset associated with larger values of $k$ happens later those of smaller $k$. However, if $f<\varepsilon$, the onset of larger $k$ occurs quicker than those of smaller $k$ (Fig. \ref{fig1}C).\\

\begin{figure}
\centering
\includegraphics[width=0.99\linewidth]{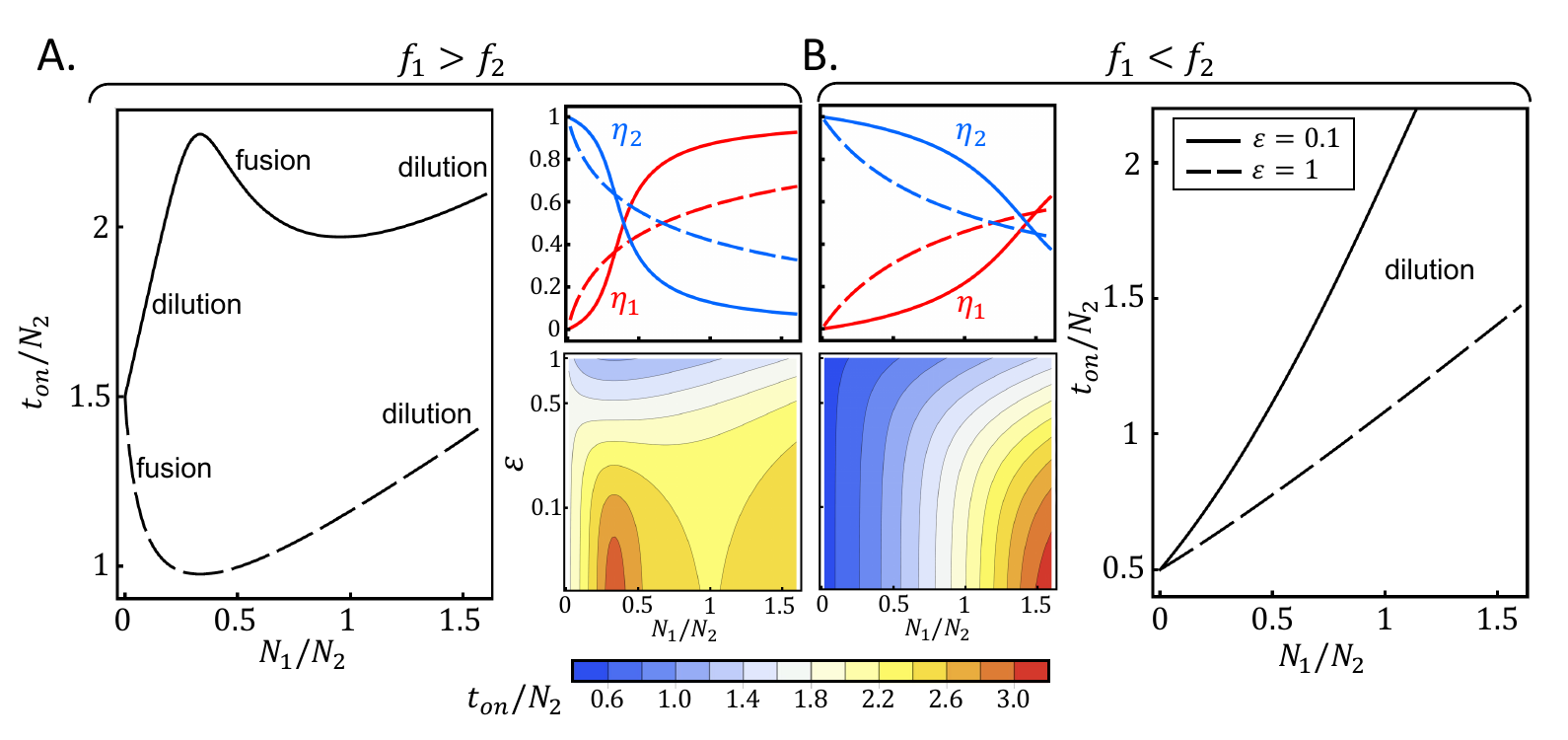}
\caption{\small{{\bf Fusion mechanisms in the two-species system.} GCC onset time $t_{on}$ and occupancy $\eta_j$ for a 2-species system as a function of the cross-species interaction $\varepsilon$, and the species concentration ratio $N_1/N_2$ ($N_2$ is kept fixed at $500$ objects). Each panel shows a large plot of the onset time for the cases of weak (solid curve) and strong (dashed curve) $\varepsilon$. In addition, two smaller plots indicate the GCC composition for each species (red and blue are species 1 and 2, respectively) these two values of $\varepsilon$ (top plot), and the full landscape of the onset time as a function of $\varepsilon$ and the concentration ratio (bottom plot). A. $f_1>f_2$ ($f_1=1$, $f_2=1/3$). B. $f_1<f_2$ ( $f_1=2/3$, $f_2=1$). In all cases the solid curve indicates $\varepsilon=0.1$, and the dashed curve $\varepsilon=1$.}}
\label{fig2}
\end{figure}

\noindent The second moment also informs about the GCC occupancy associated with each species. We define $\eta_{i}(t)$ as the fraction of species $i$ objects present in the GCC at time $t$. A good approximation of this quantity at the onset is (see Methods):
\begin{equation}
\eta_{i}(t_{on})\approx\frac{\sqrt{M_{ii}^{(2)}(t)}}{\sum_{j=1}^{k}\sqrt{M_{jj}^{(2)}(t)}}\Bigg\rvert_{t\rightarrow t_{on}}
\label{eq:eta}
\end{equation}
\noindent where $M_{ij}^{(2)}(t)$ is the matrix element of the second moment tensor at time $t$. Equations \ref{eq:roots} and \ref{eq:eta} uncover a rich micro-scale competition for the simplest multi-species system, i.e., $k=2$. Increasing the number of objects adds a dilution effect to the system that tends to delay the GCC onset. However, if the added objects are highly reactive, they would fuse quicker, which could compensate for the dilution delay and the onset could occur rather sooner. We unpack the regimes where these mechanisms dominate by varying the concentration of species 1 for two parameter regimes: $f_1>f_2$ and $f_1<f_2$. Figure \ref{fig2} illustrates this for different parameters. If $f_1>f_2$ (Fig. \ref{fig2}A), the value of $\varepsilon$ yield two main regimes of behavior with opposite effects in the GCC onset. For strong $\varepsilon$, the formation of mixed clusters occurs quicker effectively reducing the system's GCC onset in spite of the effect due to dilution. However, as the concentration grows, the dilution effect becomes greater than the fusion rate and the onset is pushed to later times. Contrarily, a weak $\varepsilon$ does not produce quicker fusion events, but it rather harms the formation of mixed clusters. Accordingly, the dilution produced by the species 1 clusters becomes dominant yielding a delay in the GCC onset. Such GCC is comprised mostly by species 2 as shown at the top right panel of Fig. \ref{fig2}A. For greater species 1 concentration values, a connected component comprised mostly by species 1 gains dominance given its larger same-species interaction value and it eventually becomes the GCC of the system causing an onset time drop. When species 1 reaches a concentration equal to species 2 and greater, the dilution effect also grows and pushes the GCC onset to later times. By contrast, if $f_1\leq f_2$ (Fig. \ref{fig2}B), the dilution mechanism is dominant regardless of the value of $\varepsilon$. As $N_1$ grows, the GCC onset occurs at later times and it is dominated by species 2 objects as shown at the top left panel of Fig. \ref{fig2}B. The full spectrum of $\varepsilon$ detailing the transition between these regimes of behaviors are shown in the bottom central panels of Fig. \ref{fig2}. 

\section{Fibrils inhibition within a biomolecular condensate}
\begin{figure}
\centering
\includegraphics[width=0.85\linewidth]{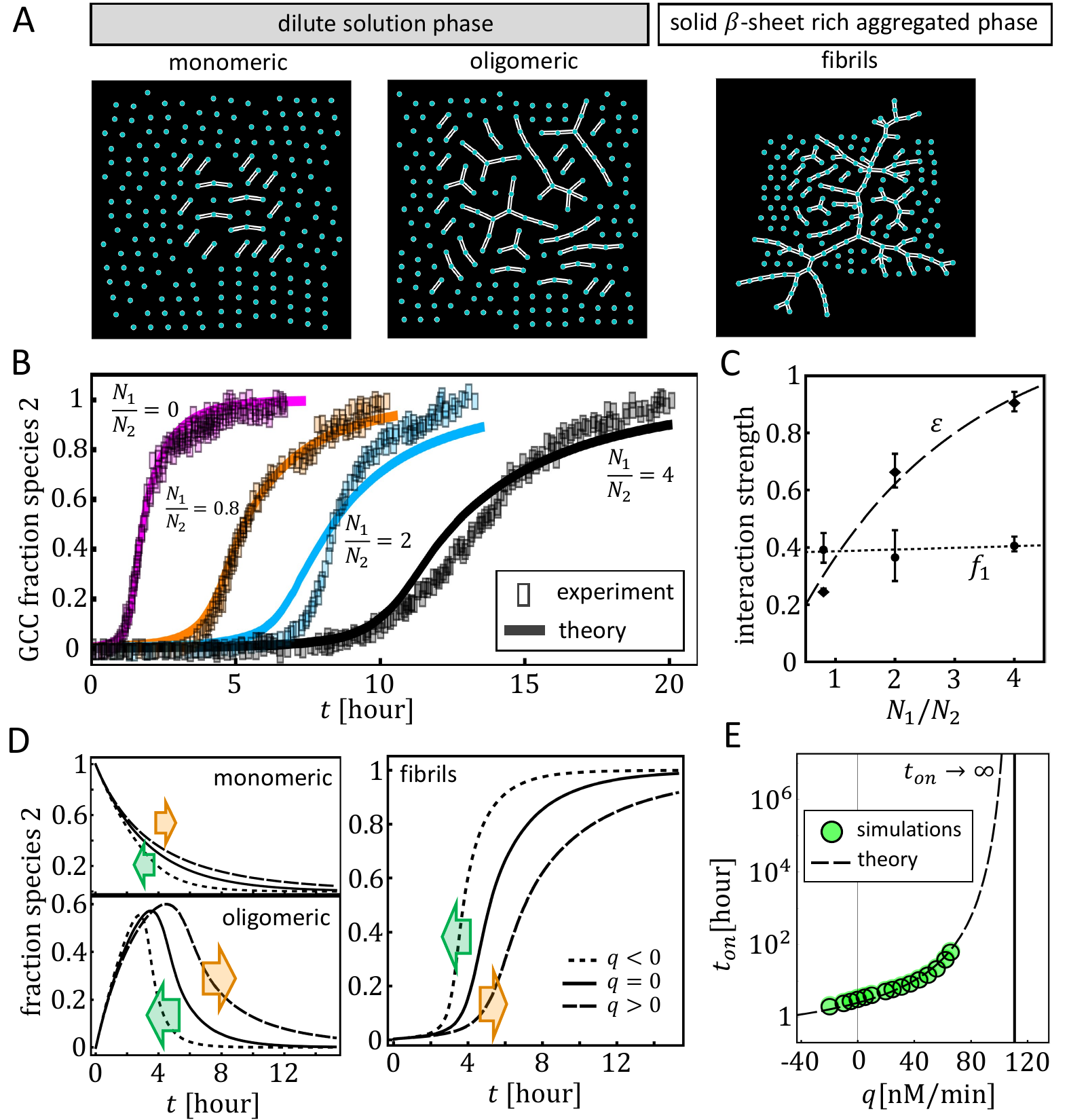}
\caption{\small{{\bf Fibrils formation in a biomolecular condensate}. A. Schematics of the dynamical phases of the amyloid peptide aggregation process. Starting from a dilute solution of monomeric peptides (left panel), dimers, trimers and higher order oligomers form through a first nucleation process yielding an oligomeric state (central panel). Resultant oligomers fuse with each other through a second nucleation yielding the solid $\beta$-rich aggregated phase akin to the emergence of the GCC in a dynamical network. B. Experimental aggregation profiles (squares) of 3.5$\mu$M A$\beta$-42 peptide solution in different concentrations of LCD Dpb1N-AK-Dpb1N as reported in ref. \cite{arosio21}, and our two-species theory (curves). The LCD concentrations are: $0 \mu$M (magenta), $2 \mu$M (orange), $5 \mu$M (blue), and $10 \mu$M (black). Accordingly, the theory uses values of concentration of species 1 (i.e., biomolecular condensate) following identical proportions (respectively, $N_1=0,400,1000,2000$), while species 2 (i.e., the amyloid system) is kept constant ($N_2=500$ objects). C. Fitting parameters ($\varepsilon$ and $f_1$) for the different concentrations. Dashed and dotted lines are exponential and linear fits, respectively. Error bars are the associated standard errors. D. Impact of dynamical changes in the concentration of the condensates in the phases of the fibril formation. The concentration increases/decreases at a constant rate $q$ ($q\approx -10,0,+10$ nM/min). Arrows point the direction the amyloid sub-system takes due to the perturbation: orange is $q>0$, and green is $q<0$. E. Theoretical prediction of the impact of dynamical variations in the condensates' concentration on the fibril onset. If the concentration growth rate reaches a threshold $q*\geq 110$nM/min, the fibril will not be formed.}}
\label{fig3}
\end{figure}

Protein assembly underlies large-scale biological processes ranging from  beneficial natural functions \cite{Maji09,buehler11,Knowles16} to aberrant disease developments \cite{soto18,hipp19,fuxreiter22}. The self-assembling behavior of amyloidogenic proteins yield the formation of amyloid fibrils, which are insoluble $\beta$-rich sheet protein structures associated with neurodegenerative disorders such as amyotrophic lateral sclerosis, Alzheimer and Parkinson's diseases \cite{chen17, irwin13,Ghulam14}, and pathologies like type 2 diabetes \cite{Butler03,Ghulam14}. The formation process of amyloid fibrils covers two dynamical phases akin to a gas-solid phase transition \cite{knowles23}. Starting from a dilute solution phase, monomeric peptides self-assemble through nucleation yielding an oligomeric intermediate state (schematics in Fig. \ref{fig3}A). Resultant oligomers grow by aggregating to monomers of the solution and by fusing with other oligomers yielding the emergence of ordered solid-like amyloid fibrils through a second nucleation process \cite{Knowles16,Arosio16,Tessa22,knowles23}. The characteristics of the resultant structure are largely independent of the folding patterns of the monomers in the fibrils and their amino acid sequence, implying that general physical principles, including a universal aggregation kinetics framework, will be adequate for studying the emergence of amyloid fibrils \cite{knowles23}.\\

\noindent Protein assembly is regulated by the local interactions, concentrations, and the properties of the surrounding cellular environment. During the past decade, researchers have discovered that the cytosol, considered as a homogeneous intracellular environment, enjoys a rich organization in compartments known as biomolecular condensates, which are able to hold biopolymers at high local concentrations in the absence of an encapsulating membrane \cite{wang21,banani17,sabari20}. Hence, the presence of intracellular condensates could drastically impact the aggregation process of amyloidogenic proteins \cite{banani17,spruijt19,spruijt22}. Experiments on the amyloidogenic peptide Abeta42 (A$\beta$-42) in biomolecular condensates formed by multi-domain proteins containing low complexity domains (LCD), show that the fibrils onset is delayed, instead of promoted, by the presence of the condensates to a degree that depends on the condensate's concentration \cite{arosio21} (symbols in Fig. \ref{fig3}B). Mathematical models consisting on adjustable rate equations \cite{Arosio16,weber19}, capture the global formation dynamics \cite{arosio21}. However, they do not explicitly quantify the interactions between the amyloid peptide and the biomolecular condensate. Hence, they do not prove nor disprove the hypothesis that the heterotypic interactions between the scaffold proteins and the peptides inhibit the fibril onset. They also do not answer the question about why other proteins could experience the opposite effect while in a biomolecular condensate as shown in refs. \cite{weber19,spruijt22}. \\

\noindent We now use our multi-species theory (with $k=2$) to explore these questions since it  describes the interactions and hence impact of a crowded and heterogeneous environment on the aggregation of recruited proteins with very different dynamical properties. Our theory accounts for first and second nucleation processes, as well as elongation, since clusters can grow by merging with monomers or larger clusters (i.e., oligomers). Starting from a monomeric state, the theory describes the dynamical transition between the dilute solution phase and the fibril phase showing a good agreement with the experiments (curves in Fig. \ref{fig3}B), where species 1 represents the LCD biomolecular condensate and species 2 the amyloidogenic peptides. The multi-domain composition of the biomolecular condensates suggests that its internal heterogeneity is greater than the
A$\beta$-42 peptides pointing to the $f_1>f_2$ regime. Thus, we set $f_2=1$ and $f_1$ is estimated from the data. The first calculation in the absence of the condensates (magenta curve in Fig. \ref{fig3}B) is a single species system. The subsequent calculations gradually increase the concentration of species 1 according to the experiments. In the second calculation (orange) the concentration is 40\% of that of the third calculation (cyan), which in turn, is 50\% of that of the fourth calculation (black). For each case, we perform a nonlinear fit to estimate the theoretical parameters $f_1$ and $\varepsilon$. The results are shown in Fig. \ref{fig3}C. The goodness-of-fit, measured by its $r^2$ value, is greater than 0.97 in all cases and we find consistent trends for each variable. The values of $f_1$ remain stable regardless of the concentration changes, and smaller than $f_2$, validating our initial assumption about the greater heterogeneity of the biomolecular condensate. Interestingly, $\varepsilon$ follows a nonlinear monotonic growth as the species 1 concentration increases, uncovering a nonlinearity in the interaction between the scaffold proteins of the biomolecular condensate and the amylogenic peptides.\\

\noindent These results support the hypothesis that the heterotypic interactions between the scaffold proteins in the condensate and the amyloidogenic peptides inhibit the fibrils' onset, and provide a quantitative basis to test intervention schemes aimed at controlling the dynamics leading to the phase transition. From a theoretical perspective, since $f_1<f_2$, the dominant competing mechanism is dilution. However, our analysis indicates that increments in the concentration of LCD of the condensate augments the fusion rate with A$\beta$-42 serving as a restricting control of the dilution mechanism. Moreover, if the properties of the biomolecular condensate are such that $f_1>f_2$, our theory predicts that the emergence of the fibril from the amyloidogenic peptides could drastically accelerate, if the interaction with the scaffold proteins is weak. This could occur if a high concentration of amyloidogenic peptides is at the cytosol-condensate interface where additional aggregation pathways can be made available \cite{spruijt22}. Going further, if we alter the monomeric phase of the condensate, we could dynamically control the dilute phase duration and hence the fibril onset. We explore this scheme in Fig. \ref{fig3}D by analyzing the impact of dynamically increasing/decreasing the condensate concentration and examine the impact in the peptides. Our theory accounts for this in the third term of equation \ref{eq:multispecies}, for an input rate of $\dot{N}_1(t)=\pm q$. We explicitly quantify the acceleration and delay of the fibrils phase by dynamically altering the concentration of the biomolecular condensate. Finally, our theory predicts that the fibrils onset can be prevented, if the concentration growth rate of the condensates reaches a finite threshold, where the onset diverges (Fig. \ref{fig3}E). We estimate this singularity occurring at a rate of $q\geq 110$nM/min (see Methods). The multi-species theory becomes an alternative framework to study the assembly dynamics of protein systems while inside complex biological environments, which could have broader applications in other biosystems where the environmental heterogeneity plays a critical role in the system's behavior and evolution (e.g., tumor growth).

\section{Growth of vaccines distrust in social networks}
\begin{figure}
\centering
\includegraphics[width=0.99\linewidth]{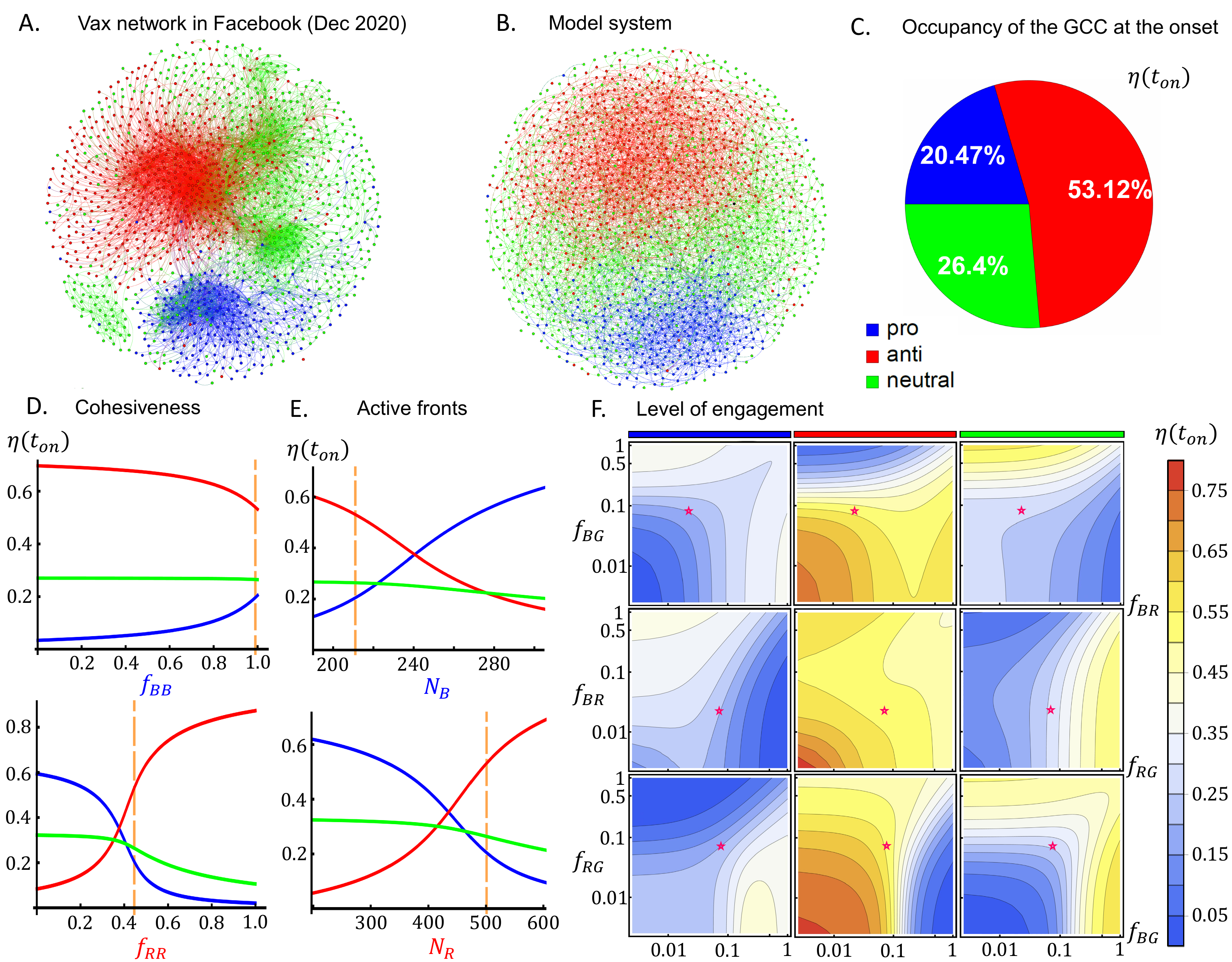}
\caption{\small{{\bf Emergence of vaccines distrust in online networks.} A. Online vaccination views network according to public Facebook pages identified in December 2020. The color of the nodes indicates the views of the specific page: 'pro' (blue), 'anti' (red), and 'neutral' (green). B. Model system constructed with our 3 species theory with parameters estimated from the real network system in A and using the same color code. C. GCC composition of the model system at the onset time ($\eta(t_{on})$) showing the dominance of the 'anti' category from the early stage. D. Impact in the GCC composition at the onset by altering the cohesiveness (i.e., $f_{ii}$) of the pro (top panel) and the anti (bottom panel). Orange vertical line in all panels of D and E point to the unperturbed parameter value for each case. F. Impact in the GCC composition at the onset by altering the number of active fronts (i.e., $N_{j}$) of the pro (top panel) and the anti (bottom panel). F. Impact in the GCC composition at the onset by altering the level of engagement for all sides of the debate: pro (left), anti (center), and neutral (right), as a function of different sets of interaction parameters: $f_{BR}$ vs. $f_{BG}$ (top row), $f_{BR}$ vs. $f_{RG}$ (central row), $f_{RG}$ vs. $f_{BG}$ (bottom row). The star on each panel points to the real values of $f_{BR}$, $f_{BG}$, and $f_{RG}$ estimated from the real data in A as shown in Table 1}}
\label{fig4}
\end{figure}

Trust in vaccines is in decline. Americans believing vaccines are safe dropped from 77\% in April 2021 to 71\% in October 2023. By contrast, those who distrust rose from 9\% to 16\% during the same period \cite{Upen23}. Besides concerns about side effects, COVID-19 unvaccinated adults reported that the top reason for being unvaccinated is distrust \cite{bureau21}. A key question is how these sentiments have spread across the population and which strategies could affect such spreading. The internet offers a path to explore these questions since it reaches more than half of the global population and it is widely used to find information \cite{kemp22}. We study a detailed community-level dataset from Facebook, the largest social media platform with 3 billion active users \cite{kemp22}. Online communities aggregate users around a common topic of interest, provide content that is publicly visible and does not require access personal information. The data collection follows a published snowball-like methodology \cite{johnson2020online,lucia22} identifying communities sharing content related to vaccines. Communities can be interconnected via a recommendation tool, and hence, a network of communities can be constructed. 1356 communities across the platform, comprising 86.7 million users, were identified to have strong content related to vaccines. Based on the content shared, 211 are classified as pro-vaccines (13 million users), 501 as anti-vaccines (7.5 million users), and the remaining 644 clusters (66.2 million users), though containing vaccine discourse, the posts are diverse in their stand and therefore their classification is neutral. From a networks perspective, this system is a complex graph of three types (i.e., species) of nodes as shown in Fig. \ref{fig4}A, which uses data collected at the heart of the COVID-19 pandemic on December of 2020. Here we refer to the pros with color blue ($B$), the antis with color red ($R$), and the neutrals with color green ($G$).\\

\noindent Previous data science studies uncovered the dominance of anti-vaccine communities in Facebook during the pre-pandemic (i.e., 2019) and pre-COVID19 vaccine emergency authorization (i.e., early 2020) periods \cite{johnson2020online,lucia22}. Despite being the smallest in terms of their number of followers, and not the largest in terms of the number of Facebook pages, anti-vaccine communities are strategically connected so their impact on neutral communities is greater than that of the pros \cite{lucia22}. To better understand what lead to this counter-intuitive outcome, we analyze this system by means of our multi-species theory (Fig. \ref{fig4}B). The estimated model parameters are in Table \ref{tab:table1} (see Methods) establishing a preliminary connection with the empirical data: (i) Pro-vaccine communities are highly cohesive and coordinated given their high $f_{BB}$ value. This is reasonable since they are run by institutions sharing methods of information dissemination. (ii) Anti-vaccines communities are more heterogeneous with $f_{RR}=0.4461$. This points to a more diverse grass-roots origin and less cohesive and coordinated. (iii) Neutral communities are the most diverse ($f_{GG}=0.1730$), which resonates with the variety of their discussion topics, e.g., parenting, healthy habits, pets. Thus, they rank the lowest in coordination and cohesiveness.\\

\begin{table}
\begin{tabular*}{0.95\textwidth}{@{\extracolsep\fill}|c||c|c|c|c|c|c|}
\hline
       $ij$  & $BB$ & $RR$ & $GG$ & $BR$ & $BG$ & $RG$\\
    \hline
    \hline
      $\lambda_{ij}$ & 1097  & 2767 & 1774 & 120 & 553 & 1152 \\
       $f_{ij}$  & 0.9999  & 0.4461 & 0.1730 & 0.0229 & 0.0821 & 0.0721 \\
\hline
\end{tabular*}
\caption{\small{Online vaccine network features and estimation of the model parameters. The number of links connecting communities of species $i$ to communities of species $j$ is $\lambda_{ij}$, with $i,j=B,R,G$, and the F-matrix elements estimated as explained in the Methods section.}}\label{tab:table1}
\end{table}

\noindent We use equation \ref{eq:roots} with $k=3$ to find the GCC onset and calculate the occupancy of each species via equation \ref{eq:eta}. The result is shown in Fig. \ref{fig4}C revealing the dominant role of the anti-vaccine movement, which is consistent with the outcome of the data science studies \cite{johnson2020online,lucia22}. This gives us confidence in the framework, and suggests that the dominance of this movement started very early in the dynamics (i.e., pre-2019). Though comprising 37\% of the communities detected, antis encompass 53.12\% of the GCC at the onset. By contrast, neutrals are more numerous with 47.5\% of the detected communities but constitute 26.4\% of the GCC at the onset. Finally, pros comprise the remaining 15.5\% of the data and have the smallest GCC occupancy with 20.47\%. This distribution is beneficial for the antis because, besides being dominant, the occupancy of their target audience is significant and their opponent is inferior.\\

\noindent We analyze the antis success and the pros failure by means of their level of cohesiveness (i.e., $f_{ii}$), number of active fronts (i.e., $N_{i}$), and cross-species engagement (i.e., $f_{ij}$, $i\neq j$). The top panel of Fig. \ref{fig4}D shows that pros would have a weaker occupancy if their cohesiveness decreases. By contrast, improving antis cohesiveness increases occupancy. However, it decreases that of the neutrals and pros alike turning the GCC into an anti-vaccines echo chamber as shown bottom panel of Fig. \ref{fig4}D. This also indicates that a decrease in cohesion of the antis simultaneously increases the GCC occupation of the pros and neutrals alike. In reference to active fronts, we find that, if the cohesiveness of the species is significant, the number of communities is closely related with the GCC occupancy at the onset (Figure \ref{fig4}E). Thus, changes in this quantity greatly benefits the pros compared to the antis given their greater level of cohesiveness. The growth rate of the GCC occupancy as a function of the number of communities added for the pros, is twice that of the antis (i.e., $\partial\eta_{B}/\partial N_{B}\approx 2\times \partial\eta_{R}/\partial N_{R}$). Similarly, the increment in the number of communities affects negatively the growth of the opponent species, and the impact against the antis is greater than that against the pros by a factor near to 5/2 (i.e., $\partial \eta_{R}/\partial N_{B}\approx (5/2)\times \partial \eta_{B}/\partial N_{R}$). Finally, the decrease in the GCC occupancy of the neutrals when more antis are added, is more than twice that of when more pros are added (i.e., $\partial \eta_{G}/\partial N_{R}\approx 2.36\times \partial \eta_{G}/\partial N_{B}$). The calculations are shown in SI sec. 2.\\

\noindent Figure \ref{fig4}F explores the impact in the GCC occupancy produced by variations of the cross-species parameters compared to the real network (star). This suggest that the pros and neutrals could increase their GCC presence, if the efforts to engage with each other increase (i.e., greater $f_{BG}$), while simultaneously suppressing the antis occupancy. Conversely, a greater engagement between antis and neutrals increases the GCC occupancy of the latter while reducing the former. This suggests that the antis have reached a critical point of dominance and influence in the GCC, where variations from this point would decrease either their occupancy or that of their target audience. Finally, interactions between pros and antis decrease the occupancy of the neutrals, however they would help gain occupancy to the side that is losing (i.e., pros). Our results shed light on the conditions leading to dominance and influence in a multi-species connected component. They suggest that the success of the antis is the result of a delicate balance between a greater number of communities than their opponent, a medium level of cohesion that does not negatively impact the occupancy of their audience, and an optimal level of engagement that allows to maintain dominance while keeping a significant audience size. 

\section{Methods}
\subsection{Theoretical Details}
\noindent For a given species $j$, we represent the unique set of phenotypes by a vector $\vec{y}^{(j)}_{i}(t)$, where $i=1,2,..,N_{k}$. Each vector entry is a real number taken from a distribution $q_j(\vec{y}^{(j)}_{i}(t))$, with $j=1,2,..., k$, that varies with the species and it also can change over time. For simplicity in the notation, here we consider one trait per object. The heterogeneous multi-species system undergo a fusion/fission dynamical process where individuals interact with each other forming and breaking clusters according to specific rules. This is quantified by the coalescence tensor $\mathcal{C}_{ij}^{(l,m)}(t)$ specifying the probability that objects $i$ and $j$ belonging to species $l$ and $m$, respectively fuse, per unit time. For example, if the cluster formation rule favors homophily between objects' traits at the individual level, the fusion probability per unit time can be defined by means of the pair's similarity $s_{ij}^{(l,m)}(t)=1-|y_{i}^{(l)}(t)-y_{j}^{(m)}(t)|$, as $\mathcal{C}_{ij}^{(l,m)}(t)=s_{ij}^{(l,m)}(t)$. Knowing the traits' distribution of the population that follows such microscopic rule, we can calculate the mean-field fusion probability at the species level 
\begin{equation}
f_{lm}(t)=\left<\mathcal{C}_{ij}^{(l,m)}(t)\right>. 
\label{eq:FM_element}
\end{equation}
Thus, for a system comprised by $k$ species, we can characterize the mean-field interactions by means of the $k\times k$ symmetric F-matrix, $\mathbf{F}(t)$, whose matrix elements are given by Eq. \ref{eq:FM_element}. 
As stated in the main text, fission processes considers a full multi-species cluster fragmentation where a cluster of size $c$ shatters into $c$ clusters of size one with a probability $\nu_{\mathrm{f}}$, which for simplicity, we set it equal across species. Additionally, we define a vector $\mathbf{s}=(s_1,...,s_k)$ to characterized a multi-species cluster, which indicates the number of objects of species $1,...,k$, respectively, that the cluster contains at a given time. Thus, if we define $n_{\mathbf{s}}(t)$ as the number of clusters whose composition is given by $\mathbf{s}$ at a given time $t$ and that clusters aggregate according to a product kernel, the fusion/fission dynamics of the system can be described by means of a further generalization of the Smoluchowski equations for heterogeneous systems \cite{char1,char2,char3,prl18,PRL2023} into a corresponding multi-species version:

\begin{eqnarray}
\dot{n}_{\mathbf{s}}&=&\frac{1}{N^2}\sum_{s_{1}'=0}^{s_{1}}...\sum_{s_{k}'=0}^{s_{k}}n_{\mathbf{s'}}\mathbf{F}\mathbf{s'}\cdot(\mathbf{s}-\mathbf{s'})n_{\mathbf{s}-\mathbf{s'}}-\frac{2n_{\mathbf{s}}}{N^2}\sum_{s_{1}'=0}^{\infty}...\sum_{s_{k}'=0}^{\infty}\mathbf{F}\mathbf{s'}\cdot\mathbf{s}n_{\mathbf{s'}}+\dot{\mathbf{N}}\sum_{i=1}^{k}\hat{e}_{i}\delta_{\mathbf{s},\hat{e}_{i}}\nonumber\\
&-&\frac{\nu_\mathrm{f}}{N}\left( n_{\mathbf{s}}\left\lVert\mathbf{s}\right\rVert_{1}\left(1-\delta_{\mathbf{s},\hat{e}_i}\right)-
\underset{\left\lVert\mathbf{s}'\right\rVert_{1}\geq 2}{\sum_{s_{1}'=0}^{\infty}...\sum_{s_{k}'=0}^{\infty}}n_{\mathbf{s}'}\left\lVert\mathbf{s'}\right\rVert_{1}(\mathbf{s}'\cdot\hat{e}_{i})\delta_{\mathbf{s},\hat{e}_i}
\right)
\label{eq:multispecies}
\end{eqnarray}

\noindent where we have omitted the time dependency to reduce the notation, $\left\lVert\mathbf{s'}\right\rVert_{1}$ is the $\ell^{1}$ norm of the vector $\mathbf{s}$, and $\hat{e}_{i}$ is the standard unitary vector associated with the $i$-th coordinate. The first two terms of equation \ref{eq:multispecies} correspond to the gain and loss fusion functions of the multi-species size distribution $n_{\mathbf{s}}$, the third describes the input/output of objects into the system, and the last describes the total fission of a multi-species cluster into individual objects. Examining the moments of the size distribution, in particular, a time singularity in the second and higher moments points to the onset of the GCC \cite{RednerBook,NewmanBook}. For a multi-species system, the second moment tensor can be derived from the vector form of an exponential generating function (see additional details in SI). For the simplest case of negligible fission and a time independent population and interactions, the second moment is 
\begin{equation}
\mathbf{M}^{(2)}(t)=\left[\mathbf{N}^{-1}-(2/N^2)\mathbf{F}t\right]^{-1}, 
\end{equation}
where $\mathbf{N}$ is a $k\times k$ diagonal matrix comprised by the number of objects $N_{j}$ of species $j$, ($j=1,2,...,k$), on each entry. Given that $\mathbf{M}^{(2)}$ is expressed as the inverse of a matrix, the singularities can be derived from the condition that determinant of the actual (i.e., non-inverted) matrix is zero, which is equation \ref{eq:roots} in the main text. \\

\noindent The second moment also informs about the GCC composition and hence it offers a quantitative route to measure which species is more or less dominant at the GCC onset. To demonstrate this aspect of the theory, we look into the matrix elements of the second moment in its traditional form as
\begin{equation}
M^{(2)}_{ij}(t)=\sum_{s_1=0}^{\infty}...\sum_{s_k=0}^{\infty}s_{i}s_{j}n_{\mathbf{s}}(t).
\label{eq:trad-m2}
\end{equation}
Focusing on the diagonal elements near the time of the GCC onset, we note that the contribution of the largest cluster dominates the sum since they grow as the square of the size. Therefore, it is a reasonable approximation of the $i$-th diagonal element near the onset to be $s_i^2$, since only one cluster dominates (i.e., for the largest cluster $n_{\mathbf{s}}=1$). Near the onset, such diagonal elements offer an approximate measure of the square of the number of species $i$ objects present in the largest component. Therefore, we could crudely estimate the proportion of objects that belong to species $i$, ($\eta_{i}$), that comprise the largest component near the onset as expressed in Eq. \ref{eq:eta}. This expression does not produce divergence issues since it is possible to separate the divergent part from the non-divergent part of equation \ref{eq:trad-m2} using equation \ref{eq:roots} (see SI sec. 1.2.4). \\

\noindent For $k=2$ and assuming that the cross-species interaction (i.e., $f_{12}\equiv\varepsilon$) is non zero, the GCC onset of the coupled system can be derived from equation \ref{eq:roots} yielding:
\begin{equation}
t_{on}=\frac{N^2}{f_{1}N_1+f_{2}N_2 +\sqrt{(f_1 N_1 -f_2 N_2)^2+4\varepsilon^2N_1 N_2}},
\label{eq:2s-ton}
\end{equation}
where $f_j\equiv f_{jj}$ is the same-species interaction of species $j$. The GCC composition is derived from the second moment yielding the proportion of objects belonging to species 1 ($\eta_1=\phi_2/\phi_0$) and 2 ($\eta_2=\phi_1/\phi_0$) at the onset, where $\phi_j$ is given as:
\begin{equation}
\phi_{j}=\left(N_{j}^{-1}-f_j\frac{2}{N^2}t_{on}\right)^{1/2},\quad j=1,2
\end{equation}
and $\phi_0=\phi_1+\phi_2$, which is general for any parameter choices. These expressions are used to calculate the results shown in Fig. \ref{fig2}. For the special case of equal concentration of objects of both species, the GCC composition acquire a symmetric behavior given by $\eta_{1(2)}=2/[\sqrt{\chi^{2}+4}-(+)\chi+2]$, where $\chi=(f_1-f_2)/\varepsilon$. Its behavior is shown in Fig. \ref{fig2}C pointing to the intuitive conclusion that the larger aggregation rate $f_j$ for one species corresponds to a larger presence in the GCC at the onset.\\

\noindent For $k=3$, the expression for the GCC onset is derived in the same way but its size is too large to be written down here. The GCC occupancy, on the other hand, is more manageable and given by $\eta_{i}=\phi_{i}/\phi_{0}$, where $\phi_{0}=\phi_{B}+\phi_{R}+\phi_{G}$, and
\begin{equation}
\phi_{i}=\left(N_{j}^{-1}N_{k}^{-1}-\left(\frac{f_k}{N_j}+\frac{f_j}{N_k}\right)\left(\frac{2t_{on}}{N^2}\right)-\left(f_j f_k -f_{jk}^2\right)\left(\frac{2t_{on}}{N^2}\right)^2\right)^{1/2},
\end{equation}
where $i,j,k=B,R,G$ and $ i\neq j \neq k$. Both quantities, the onset time and the GCC occupancy are used to calculate the results shown in Figs \ref{fig4}C-F.
\subsection{Onset divergence due to dynamical concentration increase for the two-species system}
To estimate the divergence in the onset time shown in Fig. \ref{fig3}E we employ extensive microscopic simulations of the two-species system to quantify the GCC onset. The resultant data were subsequently being used to estimate the onset divergence from the theoretical expression derived for the second moment of the single-species system with a constant input of monomers at a rate $q$ given by \cite{PRL2023}
\begin{equation}
t_{on}=\frac{N(0)}{q}\left(\left(\frac{\alpha-1}{\alpha+1}\right)^{2/\alpha}-1\right),
\end{equation}
where $4x=q(1-\alpha^2)/2$, $N(0)$ is the total population, and the divergence occurs when $q\rightarrow 8x$.
\subsection{Parameter estimation of the vax model network}
The probability per time unit that an edge connecting a community from species $i$ to a community of species $j$ is established, is given by 
\begin{equation}
f_{ij}=\lambda_{ij}/(p_{ij}\times T),
\end{equation}
where $\lambda_{ij}$ is the number of links between species $i$ communities and species $j$ communities according to the real data (Table \ref{tab:table1}), $T$ is the total time dynamics in the model, and $p_{ij}$ is the probability that two communities, belonging to species $i$ and $j$, are randomly chosen at a given time. The latter quantity is computed using the number of communities of each species $N_i$ and the total number of communities $N$: 
\begin{equation}
p_{ii}=N_{i}(N_{i}-1)/N(N-1), 
\end{equation}
for communities belonging to the same species, and 
\begin{equation}
p_{ij}=2N_{i}N_{j}/N(N-1),
\end{equation}
for communities belonging to different species, i.e., $i\neq j$. 
\subsection{Numerical simulations}
To check the robustness of our mean-field equations and test their accuracy in modeling the dynamics of cluster size distribution for a variety of interactions, we use extensive numerical simulations which incorporate the microscopic properties of our system (i.e., multi-species, individual heterogeneity). We construct the dynamical system by means of a random network approach comprised of $N(t)$ nodes where each node $j$ is uniquely identified by at time $t$ their trait $y_j(t)$ and the species $m_{j}$. Starting from a fully disconnected system, at every timestep two nodes $i$ and $j$ belonging to species $m_i$ and $m_j$ are chosen randomly from the collection. Having established at the outset of the simulation a specific set of interaction rules, with a probability $\mathcal{C}_{i,j}^{(m_i,m_j)}$ a new link is established between the nodes, and with a probability equal to $1-\mathcal{C}_{i,j}^{(m_i,m_j)}$, nothing occurs at that timestep. The process is repeated until the system is fully connected or if fission is allowed, until the dynamical steady state is reached. The accuracy of this network formation algorithm to map the dynamical kinetic equations in the mean-field limit, is demonstrated by specific examples in the SI.
\backmatter

\bmhead{Supplementary information}
Mathematical derivations, details of data collection, supplementary figures and equations that support the statements made in the manuscript.
\section*{Declarations}
\begin{itemize}
\item This research is supported by U.S. Air Force Office of Scientific Research awards FA9550-20-1-0382 and FA9550-20-1-0383. N.F.J. is also partially funded by The John Templeton Foundation.
\item The authors declare no competing interests.
\end{itemize}

\bibliography{sn-bibliography}

\end{document}